# A Coupled CFD/Trim Analysis of Coaxial Rotors[*]

Hideaki SUGAWARA,[1)†] Yasutada TANABE,[1)] and Masaharu KAMEDA[2)]

[1)] *Aviation Environmental Sustainability Innovation Hub, Japan Aerospace Exploration Agency, Tokyo 181-0015, Japan*

[2)]*Department of Mechanical Systems Engineering, Tokyo University of Agriculture and Technology, Tokyo 184-8588, Japan*

A numerical simulation method of computational fluid dynamics (CFD) coupling with a trim analysis for coaxial rotor systems is described in this paper. The trim analysis is implemented using a rotorcraft flow solver, rFlow3D. Six target forces and moments, which are the thrust of the coaxial rotor system, the rolling and pitching moments for each upper and lower rotor, and the torque balance for yaw control, are considered as the trim conditions. The blade pitch angles of both upper and lower rotors are adjusted to satisfy the target trim conditions through the trim analysis by being loosely coupled with the CFD solver. Verification of the trim analysis method and validation of the prediction accuracy of aerodynamic performance are performed based on previous experimental and numerical studies in hover and forward flight using the lift-offset conditions. It is shown that the predicted hover performances of the torque-balanced coaxial rotors are in excellent agreement with the experimental data. It is also verified that the lift-offset conditions in forward flight are simulated using this established trim analysis. Furthermore, reasonable agreements with other computational results are indicated.

*Key Words:* Coaxial Rotor, CFD, Trim Analysis, Aerodynamics

## Nomenclature

$C_D$ : drag coefficient, $D/(\rho_\infty \pi R^2 (\Omega R)^2)$

$C_{DE}$ : coefficient of effective rotor drag, $C_P/\mu + C_D$

$C_{Mx}$ : rolling moment coefficient, $M_x/(\rho_\infty \pi R^2 (\Omega R)^2 R)$

$C_{My}$ : pitching moment coefficient, $M_y/(\rho_\infty \pi R^2 (\Omega R)^2 R)$

$C_{Mz}$ : yawing moment coefficient, $M_z/(\rho_\infty \pi R^2 (\Omega R)^2 R)$

$C_P$ : power coefficient, $P/(\rho_\infty \pi R^2 (\Omega R)^3)$

$C_Q$ : torque coefficient, $Q/(\rho_\infty \pi R^2 (\Omega R)^2 R)$

$C_{relax}$ : relaxation factor

$C_T$ : thrust coefficient, $T/(\rho_\infty \pi R^2 (\Omega R)^2)$

$c$ : blade chord length, m

$D$ : rotor drag, N

$M_x$ : rotor hub rolling moment, Nm

$M_y$ : rotor hub pitching moment, Nm

$M_z$ : rotor hub yawing moment, Nm

$n$ : computational step of the trim analysis

$N_b$ : number of blades

$P$ : rotor shaft power, W

$Q$ : rotor shaft torque, Nm

$R$ : rotor radius, m

$r$ : blade or rotor radial coordinate, m

$T$ : rotor thrust, N

$z$ : Interrotor spacing, m

$\theta$ : blade pitch angle, deg

$\theta_0$ : collective pitch angle, deg

$\theta_{1c}$ : lateral cyclic pitch angle, deg

$\theta_{1s}$ : longitudinal cyclic pitch angle, deg

$\rho_\infty$ : air density, kg/m$^3$

$\sigma$ : rotor solidity, $N_b c/(\pi R)$

$\mu$ : rotor advance ratio, V/(\Omega R)

$\tau$ : rotor torque balance

$\Psi$ : azimuth angle of the blade, deg

$\Omega$ : rotor rotational speed, rad/s

Superscripts

$U$ : upper rotor of the coaxial rotor

$L$ : lower rotor of the coaxial rotor

## 1. Introduction

Several types of next-generation rotorcraft utilizing coaxial rotor systems are being developed for the improvement of high-speed capability, expanding flight range, and obtaining high hover performance at the same time. Sikorsky has demonstrated the performance potential of the Advancing Blade Concept (ABC)[1)] on XH-59A aircraft in the 1970s.[2,3)] It also developed the Sikorsky X2 Technology Demonstrator (X2TD), which was constructed using coaxial rotors and a pusher propeller, and demonstrated high-speed capability and a reduced vibration level using an active vibration control (AVC) system.[4-6)] The Sikorsky S-97 Raider is under development based on the above techniques as a light tactical prototype helicopter.[7)] NASA developed the Mars helicopter as a part of the Mars 2020 rover mission, and a coaxial rotor system was utilized.[8)] In recent Electric Vertical Take-Off and Landing (eVTOL) aircraft developments, coaxial rotor systems have been installed in several aircraft.[9)]

Numerous studies of coaxial rotors have been conducted so far. Coleman summarized the aerodynamics of coaxial rotors in hovering and forward flight in 1997.[10] Harrington experimented with two full-scale coaxial rotors under hovering conditions.[11] Both rotor blades were two-bladed and untwisted. The maximum figure of merit of the coaxial rotors was indicated to be approximately 3.3% greater than a single rotor at the same blade loading. Nagashima et al. experimentally explored wake-rotor interference effects on coaxial rotor performance when hovering.[12] Subsequently, a numerical study was performed using a generalized momentum theory and a simplified free wake analysis considering the wake-rotor interference effects.[13] The effects of wake contraction and the swirl velocity were considered in the generalized momentum theory. The numerical results of the thrust and power-sharing ratio for varying axial separation distance were correlated with the experimental data. Furthermore, Saito and Azuma computed hover performance using the Local Momentum Theory with a modified wake model for the coaxial rotor system based on the experiment conducted by Nagashima.[14] The computational results showed that the static hover performance of the coaxial rotor is predictable through comparison with the experimental data.

Recently, Ramasamy conducted experimental studies regarding the coaxial rotor. He compared the hover performances of the torque-balanced coaxial rotor, tandem rotor, and tilt-rotor configurations.[15] Torque-balanced means that the rotor torque has the same value between the upper and lower rotors in a coaxial rotor system. The influence of the axial separation distance of the coaxial rotor on aerodynamic performance was investigated for a couple of untwisted and twisted blades. The figure of merit of the coaxial rotors is improved approximately 9% compared to that of a single rotor with equivalent solidity. Cameron et al. tested a Mach-scale coaxial rotor system under hovering conditions.[16] The hover performances were compared between two- and four-bladed single rotors and two-bladed coaxial rotors. The blades were untwisted, and the blade planform was rectangular. The required power of the coaxial rotor was approximately 6% less than the four-bladed single rotor for the same thrust. Subsequently, a wind tunnel test was performed to investigate the effect of lift-offset on the rotor performance in forward flight.[17]

Analyses have been carried out based on the above experimental data to understand the aeromechanics of the coaxial rotor system and to design efficient aircraft. Leishman et al. established a blade element momentum theory (BEMT) for the coaxial rotor system when hovering and axial flight conditions, and an optimum coaxial rotor system was studied.[18] The prediction of hovering performance using BEMT correlated well with Harrington's experimental data. Similarly, several comprehensive analysis tools were validated by Lim et al.,[19] Ho et al.,[20] and Schmaus et al.[21, 22] The static aerodynamic performance of the coaxial rotor using comprehensive analysis was well predicted when compared to the experimental data. Furthermore, validations of the computational fluid dynamics (CFD) tools were conducted by Lakshminarayan et al.,[23] Singh et al.,[24] and Yoon et al.[25] The ability of CFD to predict the static hovering performance of the coaxial rotor system was indicated.

However, Passe et al.[26] and Klimchenko et al.[27] mentioned the problem of predicting the unsteady aerodynamics of the coaxial rotor using the comprehensive analysis tool. They investigated the aerodynamic interactions of the coaxial rotor in high-speed flight through CFD/Computational Structural Dynamics (CSD) coupling analysis and the comprehensive analysis tool using two-dimensional airfoil aerodynamic tables. The comprehensive analysis tool cannot reliably predict the blade sectional force variation, especially 8/rev at the blade passage frequency for a four-bladed coaxial rotor system. They suggested that high-fidelity aerodynamic analysis using CFD is required to predict the vibratory load accurately. In recent years, the importance of high-fidelity analysis such as CFD has been increasing to explore the aeromechanics of complex flowfields for rotorcraft.

Numerical simulation of the coaxial rotor system must consider the torque balance between the upper and lower rotors as a trim condition. Several CFD tools can be combined with comprehensive analysis tools to attain the trim condition of the coaxial rotor system.[24,26,27] It is difficult for the other CFD tools to carry out an analysis of the coaxial rotor system because the specific published or developed comprehensive code is required. Therefore, the trim analysis method for the coaxial rotor system is desired to be considered within the CFD tool. Lakshminarayan implemented the trimming procedure to obtain the target thrust and yaw-trimmed condition.[28] However, he only considered the system thrust and the torque balance under hovering conditions. The required blade control angles of both rotors were computed using the system of equations based on the rotor control input vector and the response vector. The derivatives of the Jacobian matrix were evaluated utilizing a vortex filament method. However, this method cannot be applied to forward flight conditions for the coaxial rotor system having a swashplate because changing cyclic pitch angles is not considered in the analysis.

The present work aims to establish an analysis technique using CFD coupling with a trim analysis for the coaxial rotor system. The proposed trim analysis is constructed to be able to simulate the coaxial rotor in both hovering and forward flight conditions, including the lift-offset state. It can be achieved simply without combining other analysis tools, such as the comprehensive analysis tools. The validations accurately predicting the rotor performance when hovering and forward flight are performed based on previous experimental and numerical studies. The trim analysis method for the coaxial rotor system and the results of primary validations are described in this paper.

This paper is composed as follows: the numerical method is described in Section 2. In Section 3, the validation results under hovering conditions are explained. Then, the computational results during forward flight with the lift-offset are described in Section 4. Finally, the conclusions of this work are summarized in Section 5.

## 2. Numerical Methods

### 2.1. Rotorcraft flow solver, rFlow3D

A rotorcraft CFD solver, rFlow3D, [29-33] developed at JAXA is utilized to predict the aerodynamic performance of the coaxial rotors. rFlow3D is a flow solver that mainly uses a moving, overlapping structured grid system. The solution can also be obtained using an unstructured grid around a complex rotorcraft fuselage.

The numerical methodologies applied to the present study are summarized in Table 1. Note that the Reynolds number is not sufficiently large for model-scale rotor simulations, so the type of boundary layer on the blade must be determined.

### 2.2. Trim analysis method for coaxial rotor system

Trim analysis for a single rotor has already been incorporated into rFlow3D. [31, 33] The trim adjustment for the single rotor is performed based on the partial derivatives of rotor thrust, rolling and pitching moments to the rotor blade controls. The partial derivatives are determined using the Blade Element Theory (BET). [40] In this paper, this method is extended to the coaxial rotor system. Six trim variables for the coaxial rotor system are defined. These are the rotor thrust of the coaxial rotor system, the rolling and pitching moments for each upper and lower rotor, and the yawing moment of the system. The blade pitch angles of the upper and lower rotors control these trim variables. The blade pitch angles are the collective pitch angles, the lateral and longitudinal cyclic pitch angles of the blade feathering motion. The feathering motions of the upper and lower rotors are described in first harmonic functions as shown below.

$$\theta^U = \theta_0^U + \theta_{1c}^U \cos\Psi + \theta_{1s}^U \sin\Psi \quad (1)$$

$$\theta^L = \theta_0^L + \theta_{1c}^L \cos\Psi + \theta_{1s}^L \sin\Psi \quad (2)$$

The system of equations is constructed using the trim variables and blade pitch angles of both rotors. The system of equations can be written in the following matrix form.

$$\mathbf{Ax} = \mathbf{b} \quad (3)$$

Where $\mathbf{A}$ is the coefficient matrix, $\mathbf{x}$ is the input vector, and $\mathbf{b}$ is the response vector. In order to converge the response vector to the target trim conditions, the response vector and the input vector are written in a delta form.

$$\mathbf{x} = \begin{bmatrix} \Delta\theta_0^U \\ \Delta\theta_{1c}^U \\ \Delta\theta_{1s}^U \\ \Delta\theta_0^L \\ \Delta\theta_{1c}^L \\ \Delta\theta_{1s}^L \end{bmatrix} \quad (4)$$

Table 1. Numerical methodologies.

| Governing equations | Three-dimensional compressible Navier-Stokes equations |
|---|---|
| Discretization method | Cell-vertex FVM (for background grid) Cell-centered FVM (for body-fitted grid) |
| Numerical flux | mSLAU (modified SLAU)[29] |
| Viscous flux | Second-order central difference scheme |
| Time integration | Explicit Runge-Kutta time integration method[34] (for background grid) Dual time-stepping/LU-SGS method[35] (for body-fitted grid) |
| High-order spatial accuracy | Fourth-order Compact MUSCL TVD (FCMT) interpolation method[36] |
| Turbulence model | Menter $k\text{-}\omega$ SST 2003 model[37,38] $\gamma - Re_\theta$ transition model[39] |
| Interpolation method | Tri-linear interpolation method |

$$\mathbf{b} = \begin{bmatrix} \Delta\sum C_T \\ \Delta C_{Mx}^U \\ \Delta C_{My}^U \\ \Delta\sum C_{Mz} \\ \Delta C_{Mx}^L \\ \Delta C_{My}^L \end{bmatrix} \quad (5)$$

Here, the input vector elements are the correction values to the blade pitch angles of each rotor in the next calculation step. The response vector elements are the difference between the target trim values and the CFD results for one rotor revolution at the current blade pitch angles of the forces and moments as shown below.

$$\Delta\sum C_T = C_{T_{target}} - \sum C_{T_{CFD}}$$

$$\Delta C_{Mx}^U = C_{Mx_{target}}^U - C_{Mx_{CFD}}^U$$

$$\Delta C_{My}^U = C_{My_{target}}^U - C_{My_{CFD}}^U \quad (6)$$

$$\Delta\sum C_{Mz} = C_{Mz_{target}} - \sum C_{Mz_{CFD}}$$

$$\Delta C_{Mx}^L = C_{Mx_{target}}^L - C_{Mx_{CFD}}^L$$

$$\Delta C_{My}^L = C_{My_{target}}^L - C_{My_{CFD}}^L$$

The coefficient matrix $\mathbf{A}$ is constructed with the partial derivatives of the trim condition variables to the blade pitch angles as shown below.

$$\mathbf{A} = \begin{bmatrix} \frac{\partial\sum C_T}{\partial\theta_0^U} & \frac{\partial\sum C_T}{\partial\theta_{1c}^U} & \frac{\partial\sum C_T}{\partial\theta_{1s}^U} & \frac{\partial\sum C_T}{\partial\theta_0^L} & \frac{\partial\sum C_T}{\partial\theta_{1c}^L} & \frac{\partial\sum C_T}{\partial\theta_{1s}^L} \\ \frac{\partial C_{Mx}^U}{\partial\theta_0^U} & \frac{\partial C_{Mx}^U}{\partial\theta_{1c}^U} & \frac{\partial C_{Mx}^U}{\partial\theta_{1s}^U} & 0 & 0 & 0 \\ \frac{\partial C_{My}^U}{\partial\theta_0^U} & \frac{\partial C_{My}^U}{\partial\theta_{1c}^U} & \frac{\partial C_{My}^U}{\partial\theta_{1s}^U} & 0 & 0 & 0 \\ \frac{\partial\sum C_{Mz}}{\partial\theta_0^U} & \frac{\partial\sum C_{Mz}}{\partial\theta_{1c}^U} & \frac{\partial\sum C_{Mz}}{\partial\theta_{1s}^U} & \frac{\partial\sum C_{Mz}}{\partial\theta_0^L} & \frac{\partial\sum C_{Mz}}{\partial\theta_{1c}^L} & \frac{\partial\sum C_{Mz}}{\partial\theta_{1s}^L} \\ 0 & 0 & 0 & \frac{\partial C_{Mx}^L}{\partial\theta_0^L} & \frac{\partial C_{Mx}^L}{\partial\theta_{1c}^L} & \frac{\partial C_{Mx}^L}{\partial\theta_{1s}^L} \\ 0 & 0 & 0 & \frac{\partial C_{My}^L}{\partial\theta_0^L} & \frac{\partial C_{My}^L}{\partial\theta_{1c}^L} & \frac{\partial C_{My}^L}{\partial\theta_{1s}^L} \end{bmatrix} \quad (7)$$

These derivatives are numerically calculated based on the BET, where the interference effects between the upper and lower rotors are not considered. The interference effect affects the rotor aerodynamic forces quantitatively, but it seems that the changes in blade pitch angles have little effect on the sensitivity. The following results of this paper demonstrate this point. The sectional aerodynamic forces can be computed using the two-dimensional aerodynamic coefficients table. Uniform inflow distribution is applied for hovering, and Pitt & Peters model[39] is adopted for forward flight to evaluate the induced velocity at each blade element in this paper.

The pitch angles of each blade are updated using the input vector of Eq. (4). Therefore, Eq. (3) is reverted to the following equation for calculating the input vector.

$$\mathbf{x} = \mathbf{A}^{-1}\mathbf{b} \qquad (8)$$

The blade pitch angles for the next step are calculated by obtaining the input vector from Eq. (8).

$$\boldsymbol{\theta}^{n+1} = \boldsymbol{\theta}^n + c_{relax}\mathbf{x} \qquad (9)$$

Where $n$ is the computational step of the trim analysis and $c_{relax}$ is a relaxation factor for improving computational stability and convergence. $\boldsymbol{\theta}$ is a vector of the blade pitch angles of the upper and lower rotors as shown below.

$$\boldsymbol{\theta} = \begin{bmatrix} \theta_0^U \\ \theta_{1c}^U \\ \theta_{1s}^U \\ \theta_0^L \\ \theta_{1c}^L \\ \theta_{1s}^L \end{bmatrix} \qquad (10)$$

The pitch angles of each blade are adjusted by being loosely coupled using CFD to achieve the trim condition. In the present study, trim analysis is performed every one revolution of the rotor. The pitch angles of each rotor are updated at the same time. The relaxation factor, $c_{relax}$, is usually set to 0.5 for this calculation. Occasionally smaller values are used as necessary to improve convergence. The value of $c_{relax}$ must be reduced if the updated blade pitch angle oscillates. In many cases, this state is the limitation of the rotor control to achieve the trim condition and is close to the stall angle in two-dimensional aerodynamic forces.

## 3. Validation of Coaxial Rotor Performance when Hovering

### 3.1. Computational conditions

The accuracy of our coupled CFD/trim method is validated for the Mach-scale coaxial rotor based on an experiment conducted by the University of Texas.[16] This experiment was performed using a model rotor. The rotor radius was 1.016 m and the blade chord length was 0.080 m. The blade was untwisted and rectangular. The blade airfoil was VR-12 with a 5% tab at the blade trailing-edge. Each rotor had two blades. The range of Reynolds number on the rotor blade based on the blade chord was $1.05 \times 10^5$ to $1.05 \times 10^6$. The specifications of the rotor are

Table 2.　Specifications of the rotor.

| | |
|---|---|
| Rotor radius | 1.016 m |
| Blade chord | 0.080 m |
| Blade twist | 0 deg |
| Solidity | 0.100 |
| Interrotor spacing | 0.140 m ($z/R$=0.138) |
| Rotor rotational direction | Upper: counterclockwise<br>Lower: clockwise |
| Preconing angle | 3.0 deg |
| Tip Mach number | 0.561 |

summarized in Table 2. More details are described in Cameron et al.[16]

Three different computational grids are used for this calculation: the outer background grid, the inner background grid, and moving/deforming blade grid. They form a moving, overlapped grid system. Figure 1 shows the grid system to simulate the coaxial rotor when hovering. Two of the background grids are Cartesian-type grids. The inner background grid is uniformly distributed, and the spatial grid resolution is $0.15c$. The outer background grid is non-uniformly distributed, except for the overlapped areas with the inner background grid. The outer background grid size in this overlapped region has a wider grid size of 0.30 $c$, two times the size of the inner background gird. The moving blade grid is body-fitted.

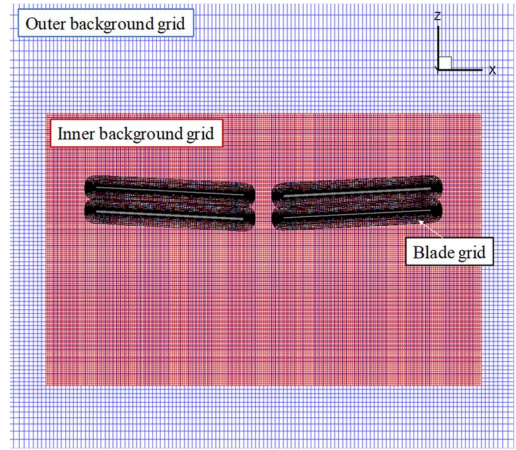

Fig. 1.　Overlapping grid system for the simulation of a model-scale coaxial rotor when hovering.

Coaxial rotor performance is computed and the blade pitch angles are changed to satisfy the trim conditions, which are the rotor thrust, the rolling and pitching moments for each rotor, and the torque balance between the upper and lower rotors. The trim conditions for hovering flight are summarized in Table 3.

The blade flapping motion is set to zero for the present computation. The blade lead-lag motion is also zero and the blade elastic deformation is not simulated. The two-dimensional aerodynamic coefficients table of a VR-12 airfoil is referred to in BET in the trim analysis. The aerodynamic coefficients are computed in the range of -180 to 180 deg using two-dimensional CFD analysis and referring to the sectional location of the blade at $r/R = 0.75$.

Table 3. Trim conditions.

| Flight condition | Hovering |
| --- | --- |
| Target $C_T (\times 10^{-3})$ | 1.75, 3.07, 4.59, 5.99, 7.05, 8.92, 9.11 |
| Target $C_{Mx}$, $C_{My}$ | 0.0 |
| Target $C_{Mz}$ | 0.0 (torque balance) |

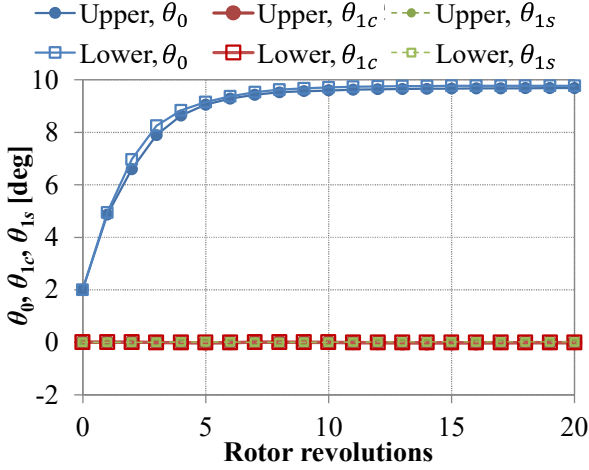

Fig. 2. Convergence histories for the blade pitch angles of the upper and lower rotors under the condition of $C_T = 8.92 \times 10^{-3}$.

Figure 2 shows the examples of convergence histories of the blade pitch angles for both upper and lower rotors under the condition of $C_T = 8.92 \times 10^{-3}$. The blade pitch angles are almost converged after 10 revolutions of the rotor. After that, the control angles of the blades are adjusted to within less than 0.1 deg.

The convergence histories of the target trim variables are shown in Fig. 3. These variables are obtained with similar convergence as those of the blade pitch angles. The error of the target rotor thrust at 10 revolutions is approximately 0.7%, as shown in Fig. 3 (a). The torque balance is approximately 0.05%, as shown in Fig. 3 (b). The torque balance is assessed as the difference of rotor torque between the upper and lower rotors normalized by the total torque of the system, as shown in the following equation.

$$\tau = \frac{C_Q^L - C_Q^U}{C_Q^L + C_Q^U} \tag{11}$$

Eventually, the errors of convergence to the target values at 20 revolutions of the rotor are 0.03% for the rotor thrust and 0.02% for the torque balance in the example case. The rolling and pitching moments of the rotor converged to the target moments (zero moments) as shown in Fig. 3 (c).

The numerical errors for all computational cases are less than 0.22% relative to the target thrust and less than 0.03% for the torque balance.

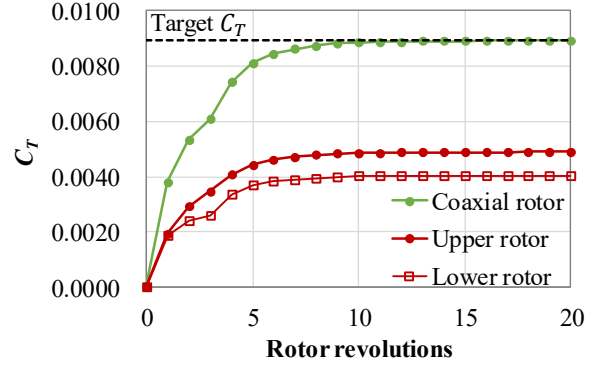

(a) Rotor thrust coefficients for each rotor

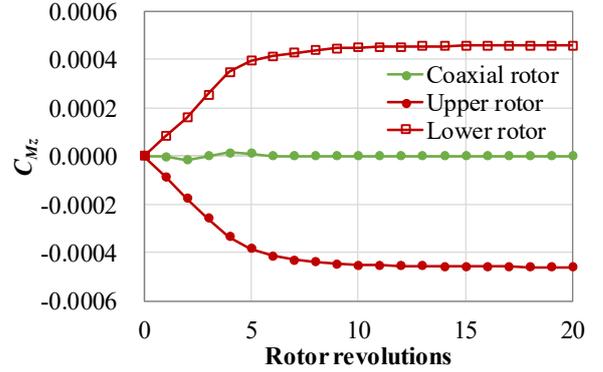

(b) Rotor yawing moment coefficients for each rotor

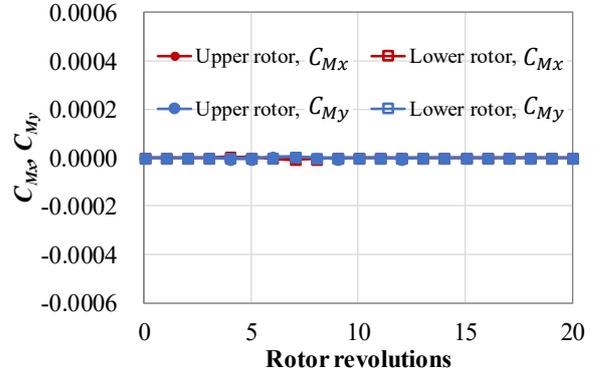

(c) Rotor rolling and pitching moment coefficients for each rotor

Fig. 3. Convergence histories of the forces and moments under the condition of $C_T = 8.92 \times 10^{-3}$.

The proposed trim method can simulate the torque-balanced coaxial rotor satisfactorily when hovering without considering the interference between the rotors in the trim analysis.

### 3.2. Computational results of aerodynamic performance for the coaxial rotor when hovering

The prediction results of coaxial rotor performance are shown in Fig. 4. The other prediction result, which was computed by Singh et al. using Helios,[41] is also indicated in Fig. 4. Singh et al. conducted the CFD/CSD simulation. The trim and CSD analysis were computed using the comprehensive analysis code RCAS. The details of the CFD method are not shown in Singh et al.[41] The predicted results of rFlow3D are in excellent agreement with the experimental data and Singh's results. Therefore, the implemented trim analysis method can satisfactorily predict the aerodynamic performance of the coaxial rotor system when hovering.

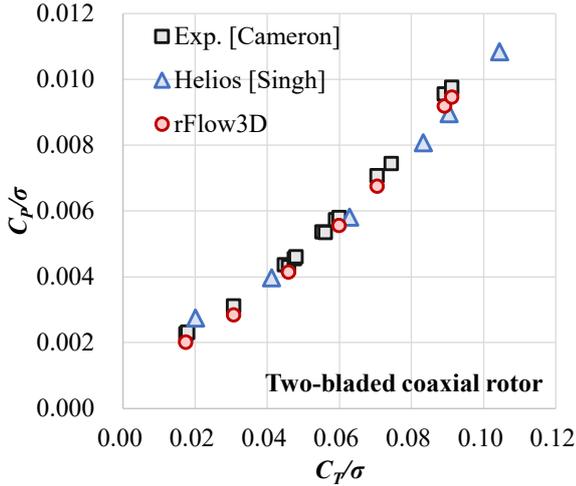

Fig. 4. Comparison of the coaxial rotor performance between the computational results and the experimental data.

The coaxial rotor has a much more complex flowfield than the single rotor, and the aerodynamic interaction occurs between the upper and lower rotors. Figure 5 visualizes the flowfield of the tip vortices under the condition of $C_T = 8.92 \times 10^{-3}$. The tip vortices from the upper rotor interact with the tip vortices from the lower rotor, after which the complex flowfield under the lower rotor can be observed.

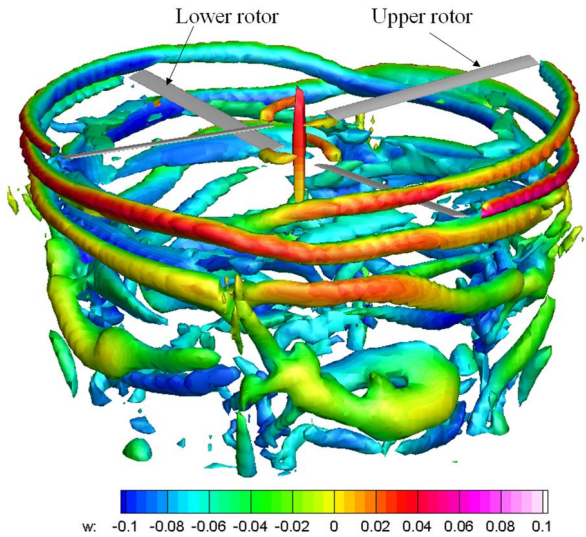

Fig. 5. Tip vortices flowfield around the coaxial rotor under the condition of $C_T = 8.92 \times 10^{-3}$. The iso-surface illustrates the tip vortices using the $Q$-criterion colored by the vertical velocity component.

Figure 6 shows the vertical velocity distribution around the coaxial rotor under the condition of $C_T = 8.92 \times 10^{-3}$. The downwash from the upper rotor interacts with the lower rotor from the half-rotor radius to the blade tip. The downwash affects the aerodynamics of the lower rotor, similar to that during climbing flight. The effective angle-of-attack of the lower rotor blade decreases more than the upper rotor at the same collective pitch angle due to the downwash of the upper rotor. The aerodynamic performance deviates between the upper and lower rotors due to the different environments of flowfield around the rotor.

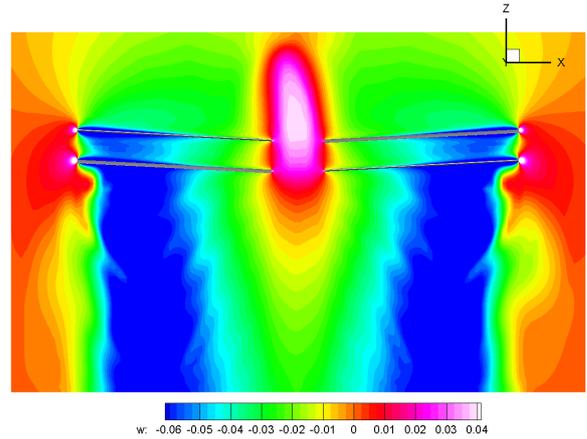

Fig. 6. Vertical velocity distribution around the coaxial rotor under the condition of $C_T = 8.92 \times 10^{-3}$.

Figure 7 shows a comparison of the aerodynamic performance for each upper and lower rotor between the experiment and the numerical results. The square symbols show the results of upper rotors and the triangle symbols indicate the results of lower rotors. The required power of the lower rotor is larger than the upper rotor at the same thrust condition, especially under high thrust conditions. The present results match well with the experimental data and Singh's results. The degradation of aerodynamic performance of the lower rotor due to the aerodynamic interaction can be simulated in excellent agreement with the experimental data.

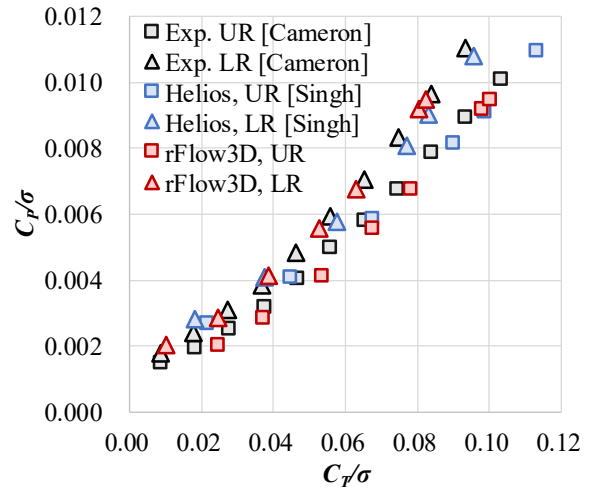

Fig. 7. Comparison of the aerodynamic performance for each upper and lower rotor between the experimental data and the numerical results. UR and LR indicate the upper rotor and the lower rotor, respectively.

## 4. Validation of Coaxial Rotor Performance during Forward Flight with Lift-Offset

### 4.1. Computational setup

The coaxial rotor during forward flight with the lift-offset is simulated to validate the implemented trim analysis method. The computational conditions are based on a previous experimental study conducted by Cameron et al.[17] The rotor specifications are similar to the previous section, as shown in Table 2. A hingeless rotor was applied in the experiment. The blade tip Mach number was 0.281, which was different from the hovering condition. The advance ratio, $\mu$, of the rotor was 0.530. The blade pitch angles were controlled to satisfy the target trim conditions. The blade flapping motion was set to zero, the blade lead-lag motion was also zero, and the blade elastic deformation was not considered here. The computational conditions are summarized in Table 4. The trim conditions are based on the measured data, which was a constant collective pitch angle of 8 deg for the upper rotor during the experiment.

Table 4. Computational conditions for model rotor during forward flight.

| Flight condition | Forward flight |
|---|---|
| Tip Mach number | 0.281 |
| Advance ratio, $\mu$ | 0.530 |
| Target $C_T (\times 10^{-3})$ | 5.95, 6.56, 7.39, 8.39, 9.56 |
| Target lift-offset | 0.01, 0.05, 0.11, 0.16, 0.22 |
| Target $C_{Mz}$ | 0.0 (torque balance) |

The lift-offset is defined concerning the rolling moment of each rotor as shown in the following equation.

$$\text{Lift-offset} = \frac{|C_{Mx}^U| + |C_{Mx}^L|}{C_T^U + C_T^L} \tag{12}$$

A similar moving, overlapped grid as in the hovering analysis is also applied, as shown in Fig. 8. The inner background grid is extended backward from the rotor center to capture the rotor wake in comparison to the grid for hovering.

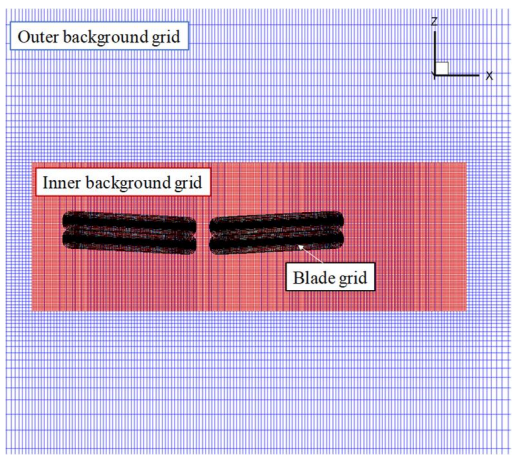

Fig. 8. Overlapping grid system for the simulation of a model-scale coaxial rotor during forward flight.

Figure 9 shows an example of convergence histories for the blade pitch angles of each rotor during forward flight with a lift-offset of 0.16. The variations of blade pitch angles are within 1.0 deg after four revolutions of the rotor. Additionally, cyclic pitch angles are required to achieve the target moment conditions of the rotors for the specified lift-offset state.

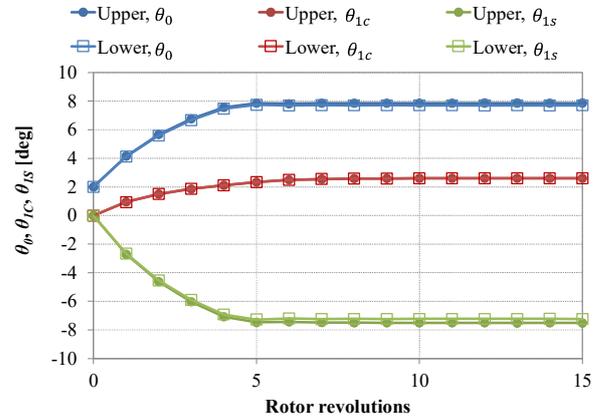

Fig. 9. Convergence histories for the blade pitch angles of the upper and lower rotors during forward flight under the lift-offset condition of 0.16.

The convergence histories of the target trim variables during forward flight with the lift-offset of 0.16 are shown in Fig. 10. These variables are almost converged after five revolutions of the rotor, the same as the histories of the blade pitch angles. The error of the coaxial rotor system thrust is less than 1% at six revolutions of the rotor as shown in Fig. 10 (a). The torque balance is less than 0.5% simultaneously, as shown in Fig. 10 (b). At 15 revolutions, the target thrust and torque balance errors are within 0.1%. The rolling moments of both upper and lower rotors arise to achieve the lift-offset state, as shown in Fig. 10 (c). The system rolling moment is canceled between both rotors. The pitching moments are also trimmed to the target conditions (zero moments), as shown in Fig. 10 (d). The implemented trim analysis can simulate the coaxial rotor satisfactorily during forward flight with the lift-offset.

### 4.2. Computational results of aerodynamic performance for forward flight with lift-offset

The computational results for forward flight with the lift-offset are described as compared to the experimental data and the previous numerical studies by Feil et al.[42] and Ho and Yeo.[43] Feil et al. investigated the aeromechanics of the model-scale coaxial rotor system with the lift-offset using a comprehensive analysis tool, CAMRAD II. Ho and Yeo utilized another comprehensive analysis tool, RCAS, and compared the rotor performance.

Figure 11 shows the vortices flowfield around the coaxial rotor during forward flight. Tip vortices are

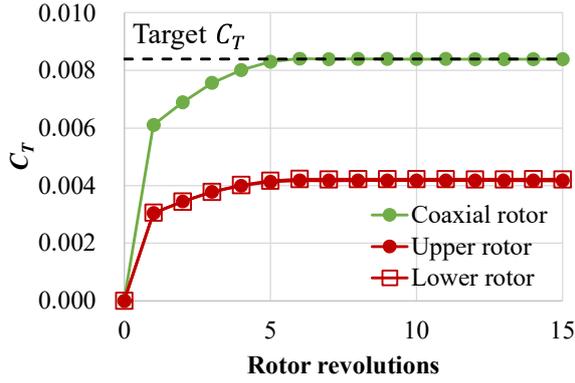

(a) Rotor thrust coefficients for each rotor

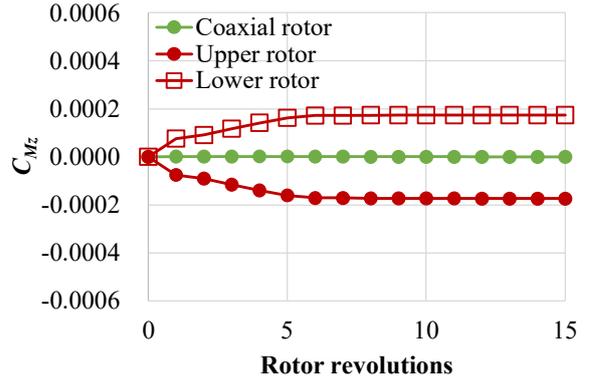

(b) Rotor yawing moment coefficients for each rotor

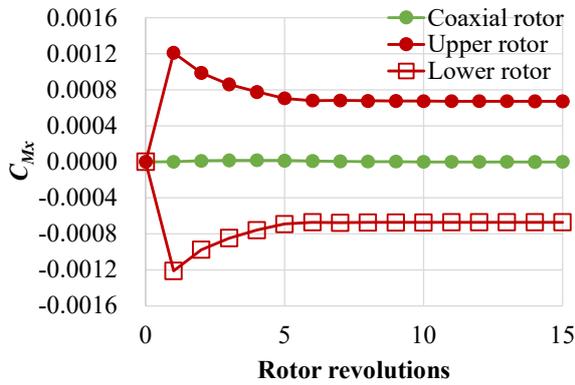

(c) Rolling moment coefficients for each rotor

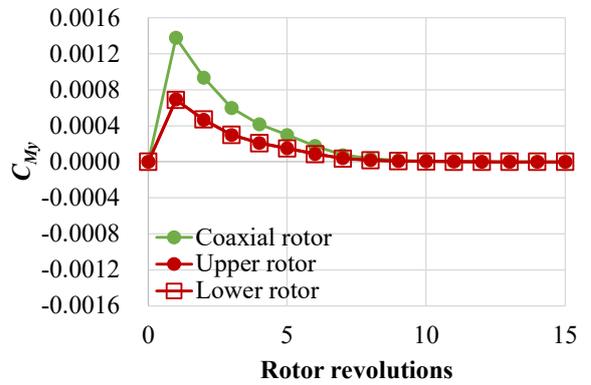

(d) Rotor pitching moment coefficients for each rotor

Fig. 10.   Convergence histories for the upper and lower rotors during forward flight with the lift-offset of 0.16.

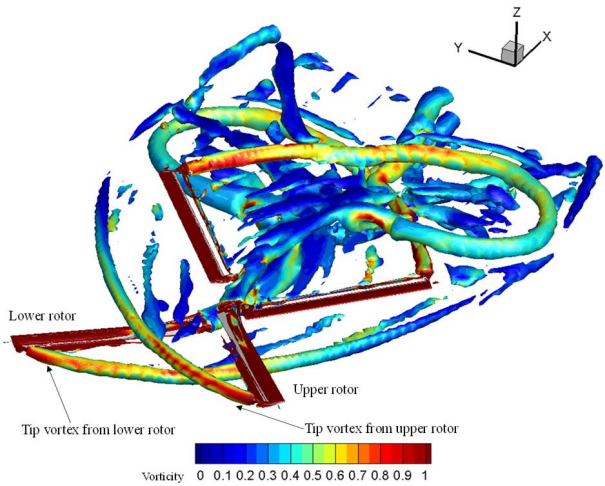

Fig. 11.   Tip vortices flowfield around the coaxial rotor under the condition of $C_T = 5.95 \times 10^{-3}$. The iso-surface illustrates the tip vortices with the $Q$-criterion colored by the vorticity.

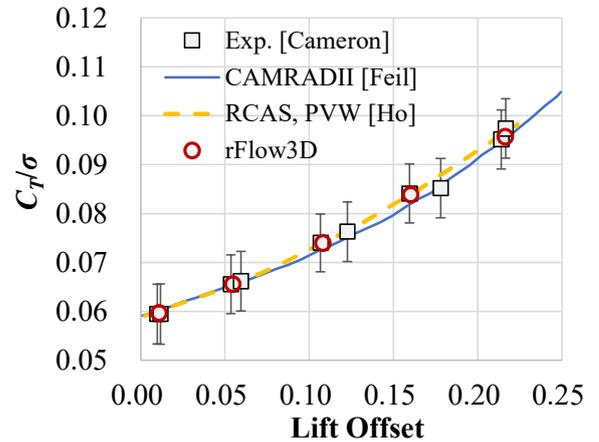

Fig. 12.   Comparison of the blade loading coefficient for the lift-offset.

generated from the upper and lower rotor blade tips, and a complex flow field is observed. Due to the high advance ratio condition, the tip vortices flow backward almost horizontally on the rotor plane.

Figure 12 shows a comparison of the blade loading coefficient against the lift-offset. The results of rFlow3D can be seen corresponding with the trend of experimental data. The predicted rotor thrust variations with respect to the lift-offset agrees with the experimental trend that the rotor thrust increases with increasing lift-offset. The other simulation results show a similar trend. The results of RCAS refer to the results using a prescribed vortex wake model (PVW).

The rolling moment coefficients of both upper and lower rotors are shown in Fig. 13. Note that the rolling moments of experimental data are not indicated Cameron et al.,[17] and these are calculated using Eq. (12). In the experiment, the rolling moments of each rotor were trimmed to achieve the target lift-offset. Figure 13 shows that the computational results of each rotor are satisfactorily trimmed to the target rolling moments using trim analysis.

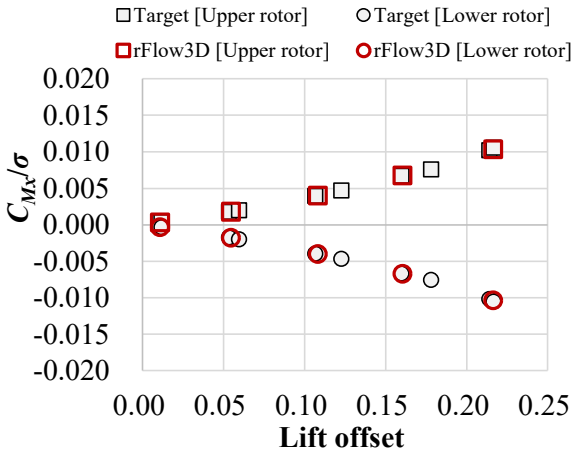

Fig. 13.    Rolling moment coefficients of upper and lower rotors for the lift-offset.

The rotor performance during forward flight is evaluated using an effective lift-to-drag ratio, $C_L/C_{DE}$. The coefficient of effective rotor drag, $C_{DE}$, consists of the drag and power of the rotor. This paper compares the drag, power, and effective lift-to-drag ratio to validate the prediction accuracy. Figure 14 shows a comparison of the rotor drag coefficient. The results using CAMRAD II are not shown because the drag of the coaxial rotor is not indicated in Feil et al.[42] The predicted results are in good agreement with the experimental data.

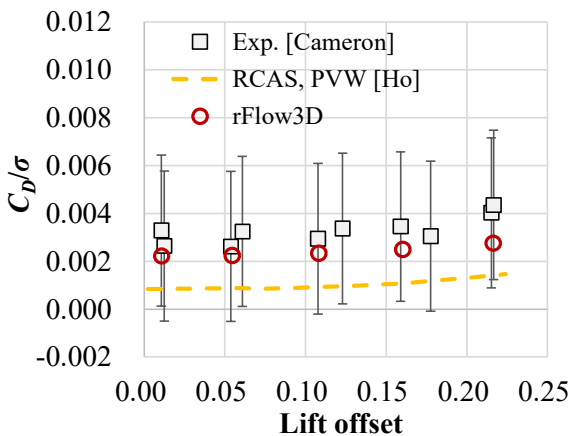

Fig. 14.    Comparison of the drag coefficients for the lift-offset.

Figure 15 shows a comparison of the rotor torque coefficients. The torque coefficient is the same as the power coefficient. The predicted rotor torque shows a different trend. The predicted rotor torque increases slightly as lift-offset increases, while the experimental results decrease. The CAMRAD II results show similar results to rFlow3D, and the results of RCAS correlate well with the experimental trend, but the results are under-predicted.

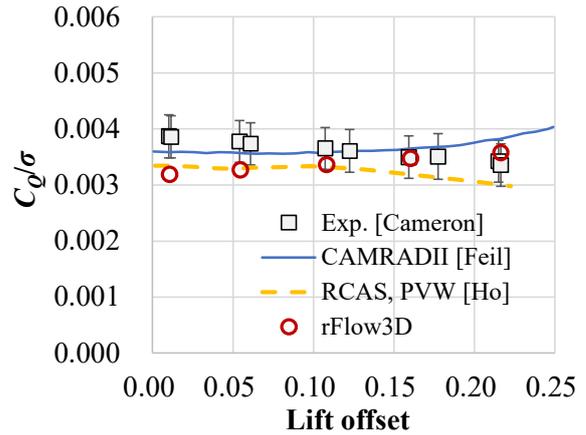

Fig. 15.    Comparison of the rotor torque coefficients for the lift-offset.

The differences in results may be due to the slight backward tilt of the rotor tip-path plane caused by elastic deformation. The lift-offset raises more lift on the blade advancing side to generate the rolling moment. Therefore, the lift-offset causes a flapping motion due to the elastic deformation, which tilts the rotor tip-path plane backward. Tilting of the rotor tip-path plane means tilting of the rotor thrust vector, which is associated with increased drag and side force of the rotor. Since the effective angle-of-attack of the blade increases with the backward tilt of the rotor tip-path plane, the required power of the rotor decreases. From Figs. 14 and 15, the rotor drag of the experiment and RCAS can be seen to increase slightly with varying lift-offset. Furthermore, the rotor torque of the experiment and RCAS both decrease as lift-off increases. Therefore, the results of rFlow3D are reasonable without considering elastic deformation. Although the difference between the two comprehensive analyses is not clear, it is probably due to the difference in elastic deformation.

A comparison of the effective lift-to-drag ratio is shown in Fig. 16. The predicted results of rFlow3D are in good agreement with experimental data and other predictions. The effective lift-to-drag ratio increases with varying lift-offset within error bars of the experimental data. It is confirmed that the coaxial rotor can also be accurately simulated during forward flight using the presently constructed trim analysis method.

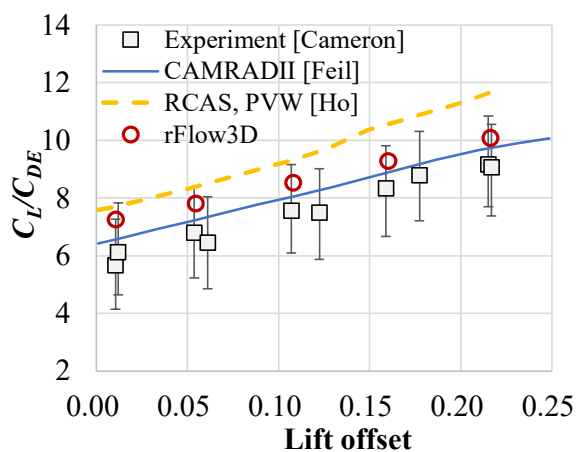

Fig. 16. Comparison of the effective lift-to-drag ratio for the lift-offset.

## 5. Conclusions

A trim analysis method coupled with computational fluid dynamics (CFD) for the coaxial rotor system is established in this paper. Six trim conditions, which are the rotor thrust of the coaxial rotor system, the rolling and pitching moments of the upper and lower rotor, and the torque balance, are defined. The trim analysis adjusts the blade pitch angles to satisfy the trim conditions using a loosely coupling analysis with CFD. The present method is implemented into an in-house rotorcraft CFD tool, rFlow3D. The validations of the proposal trim method are conducted based on previous experimental and numerical studies. The following conclusions are obtained in this paper.

1. The proposed trim analysis can simulate the torque balance and the lift-offset state of coaxial rotors together with the target thrust of the rotor system and pitching moments of the upper and lower rotors through adjusting the rotor pitch angles.
2. In calculating the partial derivatives for the trim analysis of the coaxial rotor, the sensitivity can be evaluated applying single rotor analysis using the blade element theory without considering of the interference effect between the upper and lower rotors. It is shown that smoothly converged solutions can be obtained.
3. The aerodynamic performance of the coaxial rotor system when hovering is shown to be in excellent agreement with the experimental and other numerical results. The performance change of the lower rotor due to the downwash from the upper rotor can be predicted accurately.
4. The aerodynamic performance of the coaxial rotor during forward flight with lift-offset can be predicted reasonably as proven through a comparison with the experimental and other numerical results.